\documentclass{iopart}
\usepackage{graphics}
\begin{document}
\title{Multi-spin exchange model the near melting transition of the 2D Wigner crystal}
\author{B. Bernu{\dag}\   and
D. M. Ceperley{\ddag}  }

\address{{\dag}\
  Laboratoire de Physique Th{é}orique des Liquides,
  UMR 7600 of CNRS, Universit{é} P. et M. Curie,
  boite 121, 4 Place Jussieu, 75252 Paris, France,
  \ead{bernu@lptl.jussieu.fr}
}

\address{{\ddag}\
  Dept. of Physics and NCSA, University of Illinois,
  Urbana-Champaign, Urbana, IL 61801, USA
  \ead{ceperley@uiuc.edu}
 }

\begin{abstract}
The low temperature properties of fermionic solids are governed by
spin exchanges. Near the melting transition, the spin-exchange
energy increases as well as the relative contribution of large
loops. In this paper, we check the convergence of the multi-spin
exchange model and the validity of the Thouless theory near the
melting of the Wigner crystal in two dimensions. Exchange energies
are computed using Path Integral Monte Carlo for loop sizes up to
8, at $r_s=40$,  50 and 75. The data are then fitted to a
geometric model. Then the exchange energies are extrapolated to
larger sizes in order to evaluate their contributions to the
leading term of the magnetic susceptibility and the specific heat.
These results are used to check the convergence of the  multi-spin
exchange model.
\end{abstract}

\section{Introduction}

$^3$He atoms as well as electrons carry a spin 1/2. In their solid
phases, exchange processes are responsible for the thermodynamics
at low temperature. The Thouless theory explains how a Multi Spin
Exchange (MSE) effective Hamiltonian results from these spin
exchanges\cite{thouless}. The applicability of such an approach is based on the
very different scales of the exchange energies and the phonon
energies (at least 3 orders of magnitude). In the strong coupling
limit ($r_s \to \infty$ for the Wigner crystal), a
semiclassical calculation (WKB) predicts such very small exchange
energies. Moreover, in this limit, only the two or three body
exchanges dominate depending on the dimension and the lattice
type, leading to the standard Heisenberg model (for spin 1/2,
three body exchanges are equivalent to pair exchanges). Then, at
weaker coupling, other loop exchanges arise that may lead to a
magnetic phase transition. In general, more and more exchanges are
important near the melting transition. The problem we address here
is how the MSE Hamiltonian evolves as we approach the melting
transition and whether there is a precursor to melting in the
magnetic Hamiltonian.

Exchange energies have been obtained for the bulk 3D $^3$He\cite{CJ-87} and
for $^3$He adsorbed on graphite\cite{cornell-98}. 
The exchange energies have been
obtained as well for the 2D electron Wigner crystal\cite{prl2001}-\cite{LesHouches2002}. 
For most
densities, from the semiclassical limit to near melting, the WKB
and Path Integral Monte Carlo (PIMC) evaluations assume a stable
solid of particles (a triangular lattice in 2D) without taking
into account the Fermi statistics. For these densities, only small
exchange loops with 6 or fewer particles have significant
energies. These calculations determine the exponential dependence
of exchange energies versus the density; the longer loops
decreasing more rapidly in the strong coupling limit. The MSE
model explains the main features of the magnetic transition of the
second solid layer of helium 3 adsorbed on graphite\cite{MSE}.

One problem that has been encountered near melting is that the
PIMC calculations become increasingly difficult to converge. This
is because the Boltzmannon solid melts before the fermion solid.
Adding Fermi statistics stabilizes the solid in a wider density
region. (Or alternatively, imposing antisymmetry increases the
energy of the liquid phase into which the crystal melts.) In order
to stabilize the crystal, we assume a ferromagnetic spin
polarization of ``spectator'' atoms (non-exchanging atoms) and
enforce antisymmetry on those paths. When comparison is possible,
only small changes of exchange energies are found with this
procedure. The largest differences are seen at $r_s=50$ for the
system without statistics where the solid is close to melting
(melting of the bosonic wigner crystal occurs at $r_s \approx
60$). The restriction on the path resulting from Fermi statistics
does not modify the exchange energies very much. The statistics do
not affect the local correlated motions of particles in the
vicinity of the exchanging ones. But statistics have long range
effects that stabilize the solid and forestall the melting.

The next question is whether the Thouless theory can be applied
near the melting transition. Indeed, a possible scenario could be
that the energy difference associated with a given spin-exchange,
say $J_P$,  becomes comparable to the zero-point motion kinetic
energy ($k_e^{(0)}$).  In that case, a complicated coupling
between spin states and phonon excitations could arise.
Instead our numerical results demonstrate, that all the exchange
energies are much smaller than $k_e^{(0)}$ even at melting. Even
if the exchange energy increases exponentially with the inverse
density (WKB predicts $-\log(J)\propto r_s^{1/2}$), each value
stays well below $k_e^{(0)}$. Thus, up to the melting transition,
each exchange can be seen as an independent event.

Now, near the melting transition, the number of {\sl important}
exchanges increases considerably. In this paper, we address the
question of the convergence of this MSE model versus the number of
exchanges. Indeed, if the exchange energies decrease roughly
exponentially with the loop size $n$, the number of loops of size
$n$ does increase exponentially. The convergence of the MSE model
is thus questionable. Does the total spin-exchange energy diverge
when the crystal melts? Is the existence of long-exchanges a
mechanism for the melting transition or a by-product? We try to
answer these questions from the analysis of the two-dimensional
Wigner crystal exchange data.  In section II, we present the
results of Path Integral Monte Carlo (PIMC) simulations and we
give the formula to test the convergence of the MSE model. In
section III, different models are used to fit PIMC results in
order to extrapolate to larger loop sizes.

\section{PIMC Results and Thermodynamics}
For a given permutation $P$, the exchange energy $J_P$ calculated
from PIMC needs to be monitored carefully with respect to several
parameters such as the number of particles and the time step
$\tau$. There are additional parameters associated with the
Bennett method. Fortunately, those parameters change smoothly with
the density or the type of exchange or even the system (Helium
atoms or electrons). The most important effect is the $\beta=1/T$
dependence of the results. Indeed, the calculations are done at a
temperature sufficiently low so that the phonon excitations are
non-existent but sufficiently high that the spins have not
magnetically ordered. In these conditions, the exchange energy
should be independent of the temperature. However, near melting
(the precise density depends on the statistics) results show a
linear increase versus $\beta$ (note that each independent run at
a given temperature is well converged). This is in fact
interpreted as a signature of melting that would most probably
occur for larger systems. The following tables contain only
exchange values converged and independent of $\beta$.

\begin{table}
\begin{center}
\begin{tabular}{ c|c |  r@{.}lr@{\% } |  r@{.}lr@{\% } |  r@{.}lr@{\% } }
\hline
\hline
$J$&mult& \multicolumn{3}{|c}{$r_s=$40}& \multicolumn{3}{|c}{$r_s=$50}& \multicolumn{3}{|c}{$r_s=$75}\\
\hline
$2$&3&     1402 &        & 2&      323 &        & 3&       13 & 4     & 3\\
\hline
$3$&4&     1184 &        & 3&      281 &        & 3&       14 & 2     & 2\\
\hline
$4$&6&      809 &        & 3&      165 &        & 5&        6 & 2     & 3\\
\hline
$5$&12&      295 &        & 6&       42 &        & 7&        1 & 1     & 6\\
\hline
$6_{1}$&2&      476 &        & 5&       88 &        & 9&        2 & 3     & 11\\
$6_{2}$&12&      143 &        & 7&       17 & 1     & 8&        0 & 24    & 8\\
$6_{3}$&12&       92 &        & 8&        9 &        & 9&        0 & 116   & 9\\
$6_{4}$&4&       40 &        & 13&        3 & 2     & 8&        0 & 021   & 8\\
\hline
$7_{1}$&12&      105 &        & 9&       14 & 4     & 6&        0 & 16    & 18\\
$7_{2}$&12&       72 &        & 11&        7 & 8     & 9&        0 & 076   & 21\\
$7_{3}$&24&       43 &        & 11&        3 & 6     & 9&        0 & 018   & 13\\
$7_{4}$&12&       37 &        & 21&        2 & 8     & 16&        0 & 0037  & 19\\
$7_{5}$&24&       20 &        & 15&        0 & 97    & 11&        0 & 004   & 17\\
\hline
$8_{1}$&6&      130 &        & 11&       15 & 3     & 7  & \multicolumn{3}{|c}{} \\
$8_{2}$&6&       29 &        & 13&        1 & 74    & 9  & \multicolumn{3}{|c}{} \\
$8_{3}$&12&       19 &        & 12&        1 & 19    & 10  & \multicolumn{3}{|c}{} \\
$8_{4}$&12&       23 &        & 10&        1 & 72    & 7  & \multicolumn{3}{|c}{} \\
$8_{5}$&24&       40 &        & 13&        3 & 1     & 9  & \multicolumn{3}{|c}{} \\
$8_{6}$&12&       44 &        & 12&        3 & 5     & 11  & \multicolumn{3}{|c}{} \\
$8_{7}$&6&        1 & 3     & 21&        0 & 058   & 17  & \multicolumn{3}{|c}{} \\
$8_{8}$&12&        4 & 3     & 26&        0 & 32    & 26  & \multicolumn{3}{|c}{} \\
$8_{9}$&24&        6 & 4     & 24&        0 & 25    & 21  & \multicolumn{3}{|c}{} \\
$8_{10}$&12&       23 &        & 11&        1 & 5     & 14  & \multicolumn{3}{|c}{} \\
$8_{11}$&24&        7 & 6     & 14&        0 & 25    & 20  & \multicolumn{3}{|c}{} \\
$8_{12}$&24&       18 &        & 18&        0 & 97    & 11  & \multicolumn{3}{|c}{} \\
$8_{13}$&12&        8 & 8     & 16&        0 & 24    & 15  & \multicolumn{3}{|c}{} \\
$8_{14}$&24&       22 &        & 10&        2 & 5     & 8  & \multicolumn{3}{|c}{} \\
$8_{15}$&24&        9 & 8     & 14&        0 & 51    & 10  & \multicolumn{3}{|c}{} \\
$8_{16}$&12&       14 &        & 12&        1 & 08    & 8  & \multicolumn{3}{|c}{} \\
\hline
$9_{1}$&4&       75 &        & 15&        5 & 2     & 10  & \multicolumn{3}{|c}{} \\
\hline
\hline
\end{tabular}

\caption{Exchanges energies (in $10^{-9} Ry$) obtained by PIMC
simulations for the fully polarized 2D Wigner crystal. Results are
for 64 electrons, and respectively  $\beta=1/T=$ 10000, 16600,
20000 $Ry^{-1}$, time step $\tau=$40, 66, 100 $Ry^{-1}$ for $r_s=$
40, 50, 75.
The multiplicity is the number permutation $P$ of the
same shape per site, including $P^{-1}$ symmetry.
All exchanges are for nearest neighbor on the triangular lattice.
For 2 through 6 particle exchange there is only one sort of
diagram, but for larger exchanges there are multiple possibilities
which we have labelled arbitrarily with an index. The authors can
be contacted for further details. \label{TAB:PIMCRES}
 }
\end{center}
\end{table}

Table \ref{TAB:PIMCRES} shows the PIMC results.  Shown are the
energies versus exchange for the 3 densities considered. It is
seen that exchange energies decrease as $r_s$ increase or at fixed
$r_s$, when the loop size increases. The strong dependence on the
shape of the exchange we examine next.

The overall importance of the exchange processes can be measured
by the leading term of the high temperature (HT) expansion of the
magnetic susceptibility (the Curie-Weiss temperature $\theta$) and
the specific heat:
\begin{equation}
\label{eq:Curie-Weiss}
\theta= -\sum_{n=2}^{n_{\rm max}} (-1)^n\,\frac{n(n-1)}{2^{n-1}}\, J_n^{(T)}
\end{equation}
where
\begin{equation}
\label{eq:Jtotn}
J_n^{(T)}=\sum_{shape\,\, of\,\, n\,\, sites} n_{shape} J_n^{(shape)}
\end{equation}
with $n_{shape}$ is the number of loops per site of a given
{\sl shape} and $J_n^{(shape)}$ the associated exchange value.

The leading HT-term of the specific heat is defined as
\begin{equation}
\label{eq:CVHT}
\lim_{T\to\infty} \frac{ C_V(T)}{Nk_B}\sim  \frac{9}{4}\left(\frac{J_{C_V}}{T}\right)^2
\end{equation}
where $J_{C_V}$ is given by a positive quadratic form of the exchange values:
\begin{equation}
\label{eq:JCV}
J_{C_V}^2=\sum_{ij} J_i M_{ij} J_j.
\end{equation}
Unfortunately, no simple formula exists for the matrix $\boldmath
M$. At very large $r_s$ (WKB limit) only $J_2$ and mainly $J_3$
dominate in a ferromagnetic phase and $|\theta|=J_{C_V}=J_2-2J_3$.
At $r_s\sim 200$, $J_4$ is large enough and the system becomes
anti ferromagnetic.

\section{Fits of exchange energies to analytic formula}

In order to determine the contribution of larger exchanges to the
magnetic susceptibility and specific heat in
Eqs.\ref{eq:Curie-Weiss},\ref{eq:CVHT}, we want to understand how
exchange energies vary with the loop size and shape.

Exchange energies depend mainly on the loop size $n$. Then, for a
given loop size $n$, the exchange is much larger if the exchange
contains only smooth angles, that is when its area ${\cal A}$ is
maximum (${\cal A}=(n-2+2p){\cal A}_0$, where ${\cal A}_0$ is the
area of the enclosed triangles and $p$ is the number of lattice
sites inside the loop). In order to quantify this shape dependence
we assume that each local section of the exchange contributes to
the exchange probability and use the following formula:
\begin{equation}
\label{Eq:def_alpha} \log(J_{n,s}^{fit})=\alpha_0 + \alpha_n n +
\alpha_p p_s + \sum_v \alpha_v N_{s,v}.
\end{equation}
Here $p_s$  is the number of non-exchanging sites contained inside
the loop of shape  $s$  and   $N_{s,v}$ is the number of patterns
of type $v$ encountered in the  loop of shape  $s$. The $\alpha$'s
are parameters determined from a least-squares fit minimizing:
\begin{equation}
\label{Eq:chi2} \chi^2=\frac{1}{N_J}\sum_{n,s} \left( \frac{ \log(
J_{n,s}/J_{n,s}^{fit}) }{\sigma_{n,s}} \right)^2,
\end{equation}
where $N_J$ are the number of degrees of freedom (the number of exchanges minus
the number of fitting parameters),
and $\sigma_{n,s}$ is the Monte Carlo error of $ \log(J_{n,s})$. We
now discuss the results for the fit allowing for 3 successively
more detailed patterns.

{\bf Fitting without patterns.} Without using any of the
parameters that depend on the shape of the exchange
($\alpha_v=0$), only 3 parameters remain: $\alpha_0$, $\alpha_n$
and $\alpha_p$. The results are given in table \ref{TAB:all}-Top.
Such a fit is not accurate as can be seen by the large value of
$\chi^2$. As an interpretation of the parameters, $\exp(\alpha_0)$
is the prefactor for the initiation of an exchange, $\alpha_n$ is
the action needed for adding a link onto the exchange, and
$\alpha_p$, is a term which favors non-compact exchanges.

\begin{figure}
\begin{center}
\resizebox{6cm}{6cm}{\includegraphics{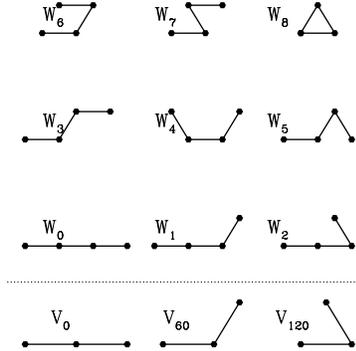}}
\caption[99]{ \label{FIG:patterns} List of the three possible
patterns made of two links (last line) and the eight patterns made
of three links.}

\end{center}
\end{figure}

{\bf Fitting with 2-links patterns.} The simplest patterns $v$
consist of the angles subtended by 2 adjacent links.  On a
triangular lattice there are three possible cases, since two
adjacent links can have angles of 0, 60, 120 or 180 degrees (see
Fig.\ref{FIG:patterns}) but the parameter $v_{180}$ only occurs in
$J_2$. Also $\alpha_n$ is becomes redundant because it can be
determined from the angles: $\sum_v N_{s,v}=n$, the number of
angles equals the number of sites in the loop. Thus we are left
with five parameters. The results for this fit are shown in table
\ref{TAB:all}-Middle.

\begin{table}
\begin{center}
\begin{tabular}{ c |  r@{.}l r|  r@{.}l r|  r@{.}l r}
\hline
\hline
$r_s$ & \multicolumn{3}{|c}{40}Ê& \multicolumn{3}{|c}{50}Ê& \multicolumn{3}{|c}{75}Ê\\
\hline
$\alpha_0$ &  8 & 99(4)&  &  8 & 34(5)&  &  6 & 79(6)& \\
$\alpha_p$ &  0 & 91(3)&  &  1 & 09(3)&  &  2 & 52(1)& \\
$\alpha_n$ & -0 & 752(9)&  & -1 & 038(9)&  & -1 & 466(14)& \\
\hline
$\chi$&\multicolumn{3}{|c}{8.5}Ê&\multicolumn{3}{|c}{9.5}Ê&\multicolumn{3}{|c}{16.0}Ê\\
\hline
\hline
\end{tabular}
\begin{tabular}{ c |  r@{.}l r|  r@{.}l r|  r@{.}l r}
\hline
\hline
$r_s$ & \multicolumn{3}{|c}{40}Ê& \multicolumn{3}{|c}{50}Ê& \multicolumn{3}{|c}{75}Ê\\
\hline
$\alpha_0$ &  12 & 1(1)&  &  12 & 32(1)&  &  12 & 9(2)& \\
$\alpha_p$ &  0 & 19(4)&  &  0 & 14(4)&  &  0 & 39(12)& \\
$\alpha_{v_{0}}$ & -0 & 883(14)&  & -1 & 185(12)&  & -1 & 954(2)& \\
$\alpha_{v_{60}}$ & -1 & 009(14)& 1.14 & -1 & 28(12)& 1.08 & -2 & 05(3)& 1.04\\
$\alpha_{v_{120}}$ & -1 & 69(3)& 1.91 & -2 & 28(3)& 1.92 & -3 & 43(7)& 1.75\\
\hline
$\chi$&\multicolumn{3}{|c}{2.5}Ê&\multicolumn{3}{|c}{4.5}Ê&\multicolumn{3}{|c}{5.0}Ê\\
\hline
\hline
\end{tabular}
\begin{tabular}{ c |  r@{.}l r|  r@{.}l r}
\hline
\hline
$r_s$ & \multicolumn{3}{|c}{40}Ê& \multicolumn{3}{|c}{50}Ê\\
\hline
$\alpha_0$ &  11 & 3(7)&  &  8 & 8(5)& \\
$\alpha_p$ &  0 & 03(8)&  & -0 & 07(7)& \\
$\alpha_{w_{0}}$ & -0 & 63(5)&  & -0 & 8(4)& \\
$\alpha_{w_{1}}$ & -0 & 76(4)& 1.20 & -0 & 8(4)& 1.0\\
$\alpha_{w_{2}}$ & -1 & 27(11)& 2.01 & -1 & 25(9)& 1.56\\
$\alpha_{w_{3}}$ & -0 & 93(4)& 1.47 & -1 & 15(3)& 1.43\\
$\alpha_{w_{4}}$ & -0 & 87(1)& 1.38 & -0 & 69(8)& 0.86\\
$\alpha_{w_{5}}$ & -1 & 37(6)& 2.17 & -1 & 65(5)& 2.06\\
$\alpha_{w_{6}}$ & -1 & 15(16)& 1.82 & -0 & 92(13)& 1.15\\
$\alpha_{w_{7}}$ & -1 & 44(7)& 2.28 & -1 & 94(5)& 2.42\\
\hline
$\chi$&\multicolumn{3}{|c}{1.3}Ê&\multicolumn{3}{|c}{1.3}Ê\\
\hline
\hline
\end{tabular}
\caption{ Top Table : Fit of exchange energies according to Eqs.
\ref{Eq:def_alpha}-\ref{Eq:chi2}. $\alpha_n$ is the main varying
parameter. Uncertainties are in parenthesis. Middle Table : Same
for fits with the 2 links patterns.  For each $r_s$, the third
column is the ratio $v_i/v_0$: within error bars, they are
independent of $r_s$ for this model. Bottom Table : same for 3
links patterns. For each $r_s$, the third column is the ratio
$w_i/w_0$, the ratio depends on $r_{s}$ for patterns with smooth
angles, and not for patterns with sharp angles. \label{TAB:npv}
\label{TAB:all}
 }
\end{center}
\end{table}

The mean error is significantly reduced by including the angles in
the fit. We find that $\alpha_0$ depends weakly on density, 
$\alpha_p$ is not very significant.
Note that the ratio $v_i/v_0$ is almost independent of $r_s$.

{\bf Fitting 3-links patterns.} In our most elaborate fit, the
possible configurations of three adjacent link are used (see Fig.
\ref{FIG:patterns}). The triangular pattern $w_8$ occurs only in
the 3-body loop, thus is removed from the fitting procedure. Note
that if the patterns for 3-links are used, those of 2-links are
redundant. Here again $\alpha_n$ is a redundant parameter. Hence,
we are left with 10 parameters for 28 exchanges for which we have
computed exchange frequencies. At $r_s=75$, we have not computed
enough different exchanges, to do this fit so we only discuss it
at smaller values of $r_s$.


The results are shown in Table \ref{TAB:all}-Bottom. This fit is
now reasonable as the mean standard deviation is of order of one,
i.e. within statistical error bars of the PIMC results. Clearly
the parameter $\alpha_p$ can be removed. In this fit, the ratios
$w_i/w_0$ depend on $r_s$ except for $w_3$, $w_5$, $w_7$.

\section{Expanding to larger loop cycles}
From the fits of the previous section, it is now possible to
evaluate, at high temperature,  the leading term $\theta$ of the
magnetic susceptibility (see Eq.\ref{eq:Curie-Weiss}) and
$J_{C_{V}}$ of the specific heat (see Eq.\ref{eq:CVHT}) by
including much larger loops. Those terms give an estimate of the
temperature of the spins begin to become correlated (note that
the expected ground state is a spin liquid). We test here if the
MSE model still converges when more and more exchanges are
involved as it appears near the melting transition.

We have calculated all self avoiding walks of length up to size $n=22$ on the triangular lattice, 
with their shape properties ($p$, angles $v$ and $w$).

\begin{figure}
\begin{center}
\resizebox{14cm}{!}{
	\includegraphics{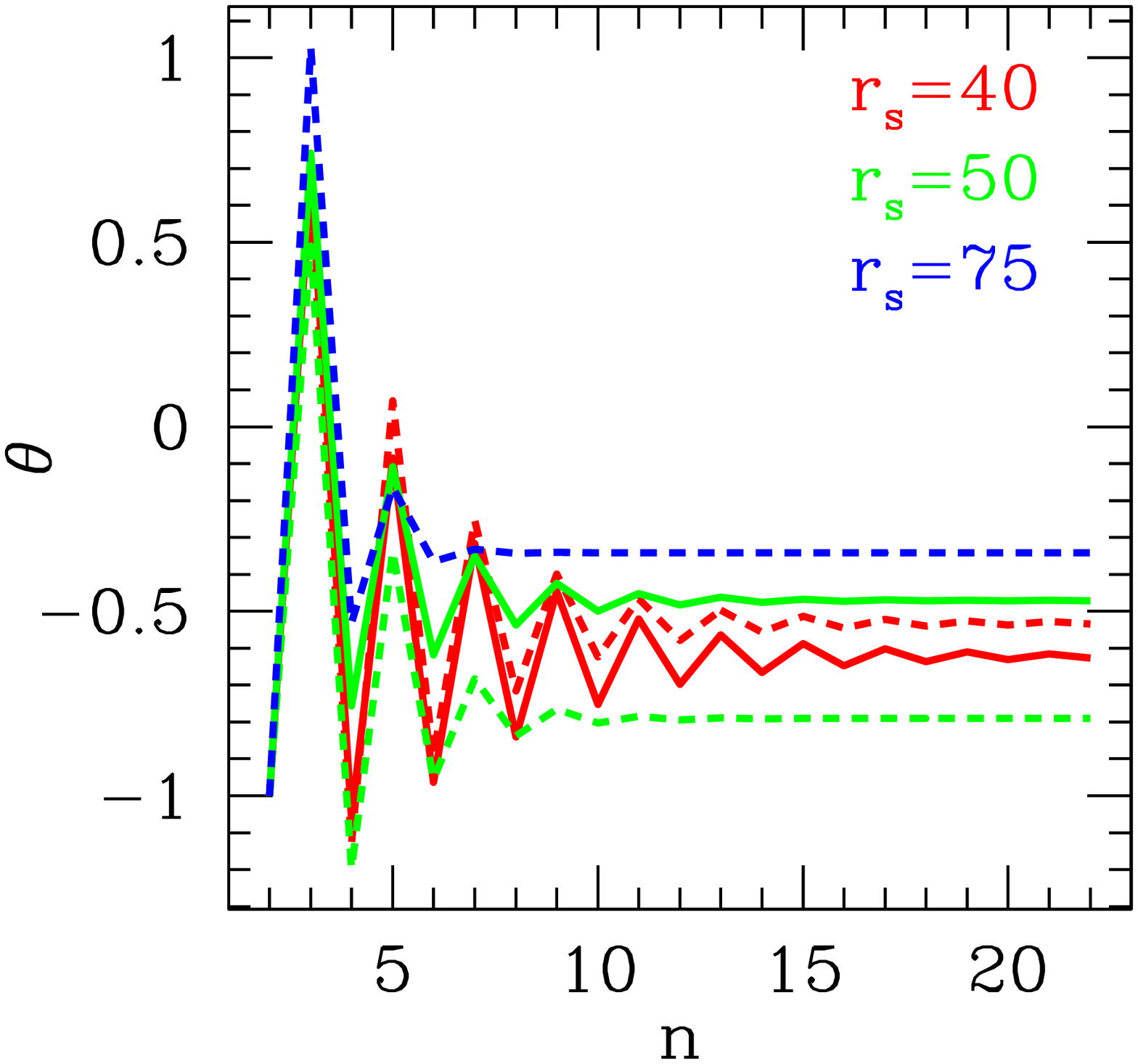}
	\includegraphics{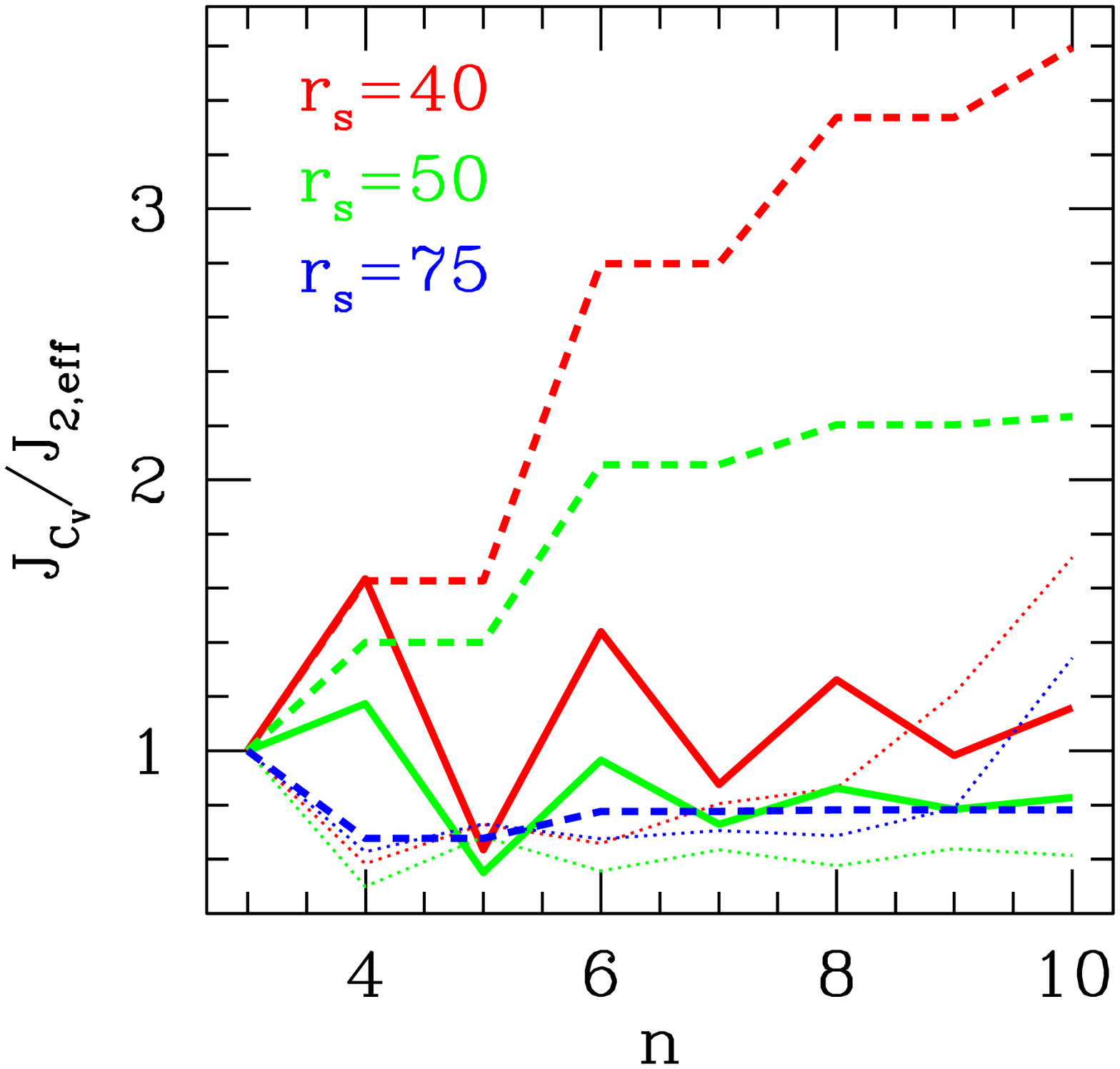}
}
\caption[99]{ \label{FIG:RESULTS} Left : Total Exchange energy of
loops of length $n$ versus $n$ (see Eq.\ref{eq:Jtotn}). Middle :
Contribution of loops of length $n$ to the Curie-Weiss temperature
versus $n$ (see Eq.\ref{eq:Curie-Weiss}). Right: Contribution of
loops of length $n$ to $J_{C_{V}}$ versus $n$ (see
Eqs.\ref{eq:CVHT}-\ref{eq:JCV}). The full lines stand for the
model with 3-link patterns, dashed lines for the model with 2-link
patterns, and the dotted line for the  model without patterns (see
text). }

\end{center}
\end{figure}

Fig.\ref{FIG:RESULTS}-left shows the sum $J^{T}_{n}$ of all
energies of loops of size $n$.  $J^{T}_{n}$ decreases at large
$r_{s}$, but clearly increases exponentially at $r_{s}=40$.
Nevertheless the Curie-Weiss temperature seems to converge at all
densities (Fig.\ref{FIG:RESULTS}-Middle) even if strong
oscillations appear at the $r_{s}=40$.

The leading term of the specific heat $J_{C_{V}}$
(Eq.\ref{eq:CVHT}) can be used to scale energies/temperatures. A
divergence of this term means a huge energy associated with
exchanges, and likely the melting of the solid.
Fig.\ref{FIG:RESULTS}-right shows a good convergence of
$J_{C_{V}}$ again at large $r_{s}$. The 2-links and 3-links models
provide rather different results. This means that $J_{C_{V}}$ is a
rather subtle combination of exchange energies. Still all models
seem to converge at $r_{s}=50$. At the smallest $r_{s}=40$, the
2-links model diverges, whereas the 3-links model seems to
converge though with strong oscillations.


\section{Conclusion}
Exchange energies can be computed with Path Integral Monte Carlo
up to the melting transition. Use of Fermi statistics is necessary
to stabilize the solid (non-local effect), but it does not change
the exchange energies (local contribution). As one approaches the
melting transition, an increasing number of different exchanges
becomes important. The Thouless theory allows one to write a
multi-spin exchange model to describe the low energy physics. But
the convergence of this model is not guaranteed and must be
verified. We have checked that for the 2D Wigner crystal, the
Curie-Weiss temperature $\theta$ and the first-order contribution
to the specific heat (Trace$(H-<H>)^{2}$) converge for the lowest
densities.
The large oscillations seen both these quantities at
$r_{s}=40$ hint at a possible divergence in the thermodynamic
properties at higher densities. Though the MSE model converges at
lower densities, a divergence of the model is possible at the
critical melting density, consistent with the picture that multi-spin
exchange is itself a mechanism for melting.
\\


\begin{thebibliography}{10}

\bibitem{thouless}
D. J. Thouless, {\it Proc. Phys. London} {\bf 86}, 893 (1965).
\bibitem{RMPI} D.~M. Ceperley, {\it Rev. Mod. Phys.} {\bf 67},
279 (1995).

\bibitem{CJ-87}
D. M. Ceperley and G. Jacucci, {\it {Phys. Rev. Lett.}}  {\bf 58}, 1648
(1987).
\bibitem{cornell-98}
B. Bernu and D.  Ceperley in {\em Quantum Monte Carlo Methods in
Physics and Chemistry},  eds. M.P. Nightingale and C.J.  Umrigar,
Kluwer (1999).

\bibitem{prl2001}
B. Bernu, L. Candido, D. Ceperley {\it Phys. Rev. Lett.} {\bf 86} 870 (2001).

\bibitem{LesHouches2002}
Bernu, B., L. Candido and D. M. Ceperley  Proceedings of the Rolduc school on Quantum Simulations of Complex Many Body Systems   (2002). 
Bernu, B. and D. M. Ceperley  J. of Physics Cond. Mat. 14, 9099 (2002),  "Exchange Frequencies in 2d solids"  

\bibitem{MSE}
G. Misguich, B. Bernu, C. Lhuillier and
C. Waldtmann. {\it {Phys. Rev. Lett.}} {\bf{81}} 1098 (1998); {\it
{Phys. Rev. B}} {\bf{60}} 1064 (1999).



\end{thebibliography}
\end{document}